\documentclass[prl,twocolumn,showpacs,preprintnumbers,amsmath,amssymb]{revtex4}
\usepackage{txfonts}
\usepackage{epsfig,amsmath,amssymb} 
\usepackage[all]{xy}
\usepackage{ifpdf} 
\usepackage{graphicx}   
\usepackage{amsfonts}
\usepackage{makeidx}  
\usepackage{epigraph}
\usepackage{etoolbox}
\makeatletter
\patchcmd{\epigraph}{\@epitext{#1}}{\itshape\@epitext{#1}}{}{}
\makeatother

\newcommand{\qed}{\hfill \mbox{\raggedright \rule{.07in}{.1in}}}

\newcommand{\ket}[1]{\left | #1 \right\rangle}

\newcommand{\qq}[1]{``#1"}

\begin{document}

\title{Gravitationally-induced entanglement between two massive particles is sufficient evidence of quantum effects in gravity} 

\author{C. Marletto$^{a}$ and V. Vedral $^{a,b}$
	\\ {\small $^{a}$ Clarendon Laboratory, Department of Physics, University of Oxford} 
	\\{\small $^{b}$Centre for Quantum Technologies, National University of Singapore}}

\date{\today}

\begin{abstract}

\noindent All existing quantum gravity proposals are extremely hard to test in practice. Quantum effects in the gravitational field are exceptionally small, unlike those in the electromagnetic field. The fundamental reason is that the gravitational coupling constant is about 43 orders of magnitude smaller than the fine structure constant, which governs light-matter interactions. For example, detecting gravitons -- the hypothetical quanta of the gravitational field predicted by certain quantum-gravity proposals -- is deemed to be practically impossible. Here we adopt a radically different, quantum-information-theoretic approach to testing quantum gravity. We propose to witness quantum-like features in the gravitational field, by probing it with two masses each in a superposition of two locations. First, we prove that any system (e.g. a field) mediating entanglement between two quantum systems must be quantum. This argument is general and does not rely on any specific dynamics. Then, we propose an experiment to detect the entanglement generated between two masses via gravitational interaction. By our argument, the degree of entanglement between the masses is a witness of the field quantisation. This experiment does not require any quantum control over gravity. It is also closer to realisation than detecting gravitons or detecting quantum gravitational vacuum fluctuations.
\bigskip

\end{abstract}

\maketitle

Contemporary physics is in a peculiar state. The most fundamental physical theories, quantum theory and general relativity, claim to be universally applicable and have been confirmed to a high accuracy in their respective domains. Yet, it is hard to merge them into a unique corpus of laws. We still do not have an uncontroversial proposal for quantum gravity. Some approaches are based on applying a quantisation procedure to the gravitational field, \cite{DEW}, in analogy with the electromagnetic field; some other are based on \qq{geometrizing} quantum physics \cite{KIB}; while others modify both into a more general theory (e.g. string theory \cite{POL}) containing both quantum physics and general relativity as special cases. All of them are affected by acute technical and conceptual difficulties \cite{KIE, WAL, SAK2}. 

There is, however, an even more serious problem. Current proposals for quantum gravity lead to seemingly untestable predictions \cite{DYS, ROT}. On this ground, some have even argued that quantising gravity is not needed after all \cite{PEN} or that gravity may not even be a fundamental force \cite{JAC,SAK}. Ronsenfeld summarised the problem as follows: \qq{the incorporation of gravitation into a general quantum theory of fields is an open problem, because the necessary empirical clues for deciding the question of the quantization of the gravitational fields are missing. It is not so much a matter here of the mathematical problem of how one should develop a quantum formalism for gravitation, but rather of the purely empirical question, whether the gravitational field - and thus also the metric - evidence quantum-like features.} \cite{ROS}.

How would one confirm experimentally that the gravitational field has \qq{quantum-like features}? A good starting point, though not sufficient, is a thought experiment Feynman proposed during the Chapel Hill conference on gravity \cite{FEY}.  A test mass is prepared in a superposition of two different locations and then interacts with the gravitational field. 
%
%

Then, the gravitational field and the mass would presumably become entangled (Feynman used different terminology, but that is what a fully quantum treatment would imply). To conclude that the field must be quantised, Feynman proposed to perform a full interference of the mass. If the mass did interfere, Feynman's reasoning goes, gravity would be quantum since re-merging the two spatial branches would then reverse the coupling to gravity, confirming the unitary dynamics in quantum theory. Of course, Feynman also acknowledged that quantum theory could stop applying at a certain scale. This would then presumably constitute a new law of nature -- for instance, see the existing \qq{gravitational collapse} literature \cite{PEN, DIOSI, KAROL}. 

Even if successful in showing the full interference of a single macroscopic mass, Feynman's thought experiment is not enough to conclude that the gravitational field is quantum. This is because his proposed interference only requires that the two spatial states of the mass acquire different phases during the experiment. These phases could simply be induced by interaction with an entirely classical gravitational field, without ever requiring entanglement between the mass and the field. There is indeed a long history of witnessing such phases induced by classical gravitational fields. Prominent examples are the Collela-Overhauser-Werner (COW) and related experiments \cite{COW}, where the phase of a single-neutron interferometer is controlled by the Earth's Newtonian gravitational potential. Another more recent proposal for a proof-of-principle experiment, involving general relativistic effects on quantum systems \cite{FOL, BRU}, relies on the gravitational redshift caused by the Earth's gravitational field affecting the phase of an interference experiment with atomic clocks. All such experiments (see also \cite{AHL}) are compatible with the gravitational field being completely classical; this is indeed the main assumption under which their predicted outcomes are derived.

So, a gravitationally induced phase on the quantum state of a single mass does not constitute experimental evidence of the quantisation of gravity. Instead, one would have to show that the gravitational field is capable of existing in a superposition of different values. The key here is to be able to witness the presence of another observable in the field that does not commute with the first one. This is precisely what one means by the field being \qq{quantum}: it must have at least two non-commuting observables. 

We show here that it is possible to witness quantum features of the field by probing it with {\sl two} masses. Intuitively, the first mass, being in a superposition of two locations, becomes entangled with the field; while the second mass, also in a superposition, is used to witness that entanglement \cite{MAVENAT}.  
This is a fundamentally different approach to detecting quantum effects in the gravitational field, based on a quantum-information-theoretic method that requires no quantum manipulation of the gravitational field itself. First, we prove the fact that if two quantum systems (e.g. two masses that can be spatially superposed) become entangled through an interaction with a third system (e.g. the gravitational field), then that third system must itself be quantum -- in the above sense of having two non-commuting observables. This argument is general in that it could apply to any system be it continuous (like a field) or discrete (like a spin). It is also independent of the exact details of the dynamics, along the lines of \cite{MAVE, MAVE1}. Interestingly, this makes our proposal independent of particular models of quantum gravity. This is an advantage given that there are many different proposals. 

Then, we propose an experiment, based on our theoretical argument, in which the two quantum systems are two masses each spatially superposed. The third mediating system is the gravitational field. Via our theoretical argument, the entanglement between the positional degrees of freedom of the masses is an indirect witness of the quantisation of the gravitational field. As we shall see, the entanglement between the masses is function of the relative phase acquired by each of the masses along the paths, via their interaction with the gravitational potential generated by the other superposed mass. This experiment, as we shall illustrate, is feasible with current technology, using some form of matter-wave interferometry -- e.g. \cite{ARN, NANO, ASE}. Our experiment only relies on having the full quantum control over the two masses. 

Consider now three physical systems: two quantum systems $Q_1$ and $Q_2$ (e.g. the two quantum masses in our experiment) and a third system $C$ (e.g. the field mediating the interaction). Suppose that $C$ is \qq{classical}, by which we mean that $C$ has only a single observable $T$. This notion of classicality is information-theoretic \cite{MAVE, MAVE1} and sharply differs from other existing ones -- e.g. the field being in a coherent state, or its being a decoherent channel \cite{MIL}. 

For simplicity, we assume that $Q_1$ and $Q_2$ are qubits. Let $\hat q^{(1)}\doteq(\sigma_x\otimes I_{2,c}, \sigma_y\otimes I_{2,c}, \sigma_z\otimes I_{2,c})$ denote the vector of generators $q_{\alpha}^{(1)}$ of the algebra of observables of the qubit ${Q_1}$, where $\sigma_{\alpha}, \alpha =x,y,z$, are the Pauli operators and $I_{2,c}$ is the unit on ${Q_2}$ and $C$.  Let $\hat q^{(2)}$ be defined in a similar way. We also assume that the classical system $C$ is a bit, i.e., $T$ is a binary observable. Without loss of generality, we can represent it as an operator $q_z^{(C)}\doteq I_{12}\otimes\sigma_z$, where $I_{12}$ is the unit on ${Q_1\oplus Q_2}$. In our proposed experiment, $T$ might be a discretised version of one of the quadratures of the gravitational field. The argument, however, would apply to continuous systems too (e.g. a harmonic oscillator). 

Now, consider an experiment where $Q_1$, $Q_2$ and $C$ are initially disentangled. For example, they are each one independently prepared in an eigenstate of $\sigma_z$.  Suppose $Q_1$ interacts with $C$ and $Q_2$ interacts with $C$, separately; but $Q_1$ and $Q_2$ never interact directly. Suppose that after these interactions $Q_1$ and $Q_2$ are confirmed to be entangled. Entanglement is confirmed by directly measuring observables on $Q_1$ and $Q_2$ only, in different basis, to implement a witness -- but no measurements are ever performed on C (despite the fact that it could be measured in its own classical basis).

That $Q_1$ and $Q_2$ are entangled is in contradiction with $C$ being classical, thus proving that it must have at least another complementary observable in addition to $T$. This is because, if $C$ is classical, the most general form of a state of $Q_i\oplus C$ is $$\rho = \frac{1}{4} \left ( I_{12,c} +\underline{r}.\hat q^{(i)}+s_z q_z^{(C)}+ \underline{t}.\hat q^{(i)} q_z^{(C)}\right)\;,$$ for some real-valued vectors $\underline r$, $\underline t$ and for some real coefficient $s_z$ ($I_{12,c}$ is the unit on $Q_1\oplus Q_2\oplus C$). This state, when interpreted as a two-qubit state, is separable. Hence, the most general state of the system $Q_1\oplus Q_2\oplus C$, if the three systems start globally disentangled and $Q_1$ and $Q_2$ can never directly interact, will also be separable. In particular, the state of $Q_1\oplus Q_2$ will be separable. Hence, if $Q_1$ and $Q_2$ are found in an entangled state and if that entanglement has been mediated by the interaction with $C$, $C$ must have itself at least another observable complementary to $T$. 

Now, a field can be considered as a collection of systems $C_i$ each one being a harmonic oscillator, mediating the interaction between two quantum systems $Q_1$ and $Q_2$ that can couple to the field. The collection of systems $C_i$ can itself be regarded as a classical system $C$. Any interaction between $Q_1$ and $Q_2$ mediated by the field can be modelled as an interaction between $Q_1$ and $C$, and then between $C$ and $Q_2$. By applying the same argument as above, if $Q_1$ and $Q_2$ can be entangled via the field, the field must be quantum in the above sense. 
Unlike other witnesses of non-classicality (e.g. \cite{PAT}), our argument does not assume any specific dynamics. The only assumption here is that the interactions must be local -- namely, there cannot be any action at a distance between $Q_1$ and $Q_2$ -- and that $Q_1$ interacts with $C$ only, and so does $Q_2$.

It is of course always possible to generate entanglement between $Q_1$ and $Q_2$ by using a classical system $C$ as the control of a controlled unitary on $Q_1 \oplus Q_2$. For example, the unitary could act so that if $C$ is in a certain state of its classical basis, a unitary prepares the system $Q_1 \oplus Q_2$ in a particular entangled state. However, generating that entangled state would require $Q_1$ and $Q_2$ to interact directly with one another, which violates our locality assumption. Note that we have, for present purposes, ignored the possibility of using non-local features of the geometry of spacetime, such as closed time-like curves \cite{CTC}. It is not excluded that by allowing such features one might be able to establish entanglement nonetheless via local interactions with $C$. Note also that it would not be possible to apply, in this context, well-known results of quantum information theory, such as the fact that Local Operations and Classical Communication cannot increase the entanglement between two spatially separated parties \cite{HOR}. This is because those results assume that all the systems involved obey quantum theory. Here, instead, the gravitational field cannot be assumed to obey quantum theory (the experiment is precisely designed to assess whether it does!).  This is why one must resort to the more general argument we propose.

We turn now to our experimental proposal -- see figure 1. Two quantum systems $Q_1$ and $Q_2$ with equal mass $m$ are entangled only via the gravitational field -- which plays the role of the system $C$. Our argument implies that the entanglement between $Q_1\oplus Q_2$ is an indirect witness of non-classicality of gravity -- i.e., of the non-commutativity of the observables on the gravitational field. 
Specifically, each mass is in one of two Mach-Zehnder interferometers, each located so that both masses are subject to the same Earth's gravitational field (for example parallel to the Earth's surface). The lower interferometer arm is indicated by $0$ and the upper arm by $1$.  Each mass is put in the state $\frac{1}{\sqrt 2} \left(\ket{0}+\ket{1}\right )$ by the first beam-splitter. 
\begin{figure}[h]
	\centering
	\includegraphics[scale=0.4]{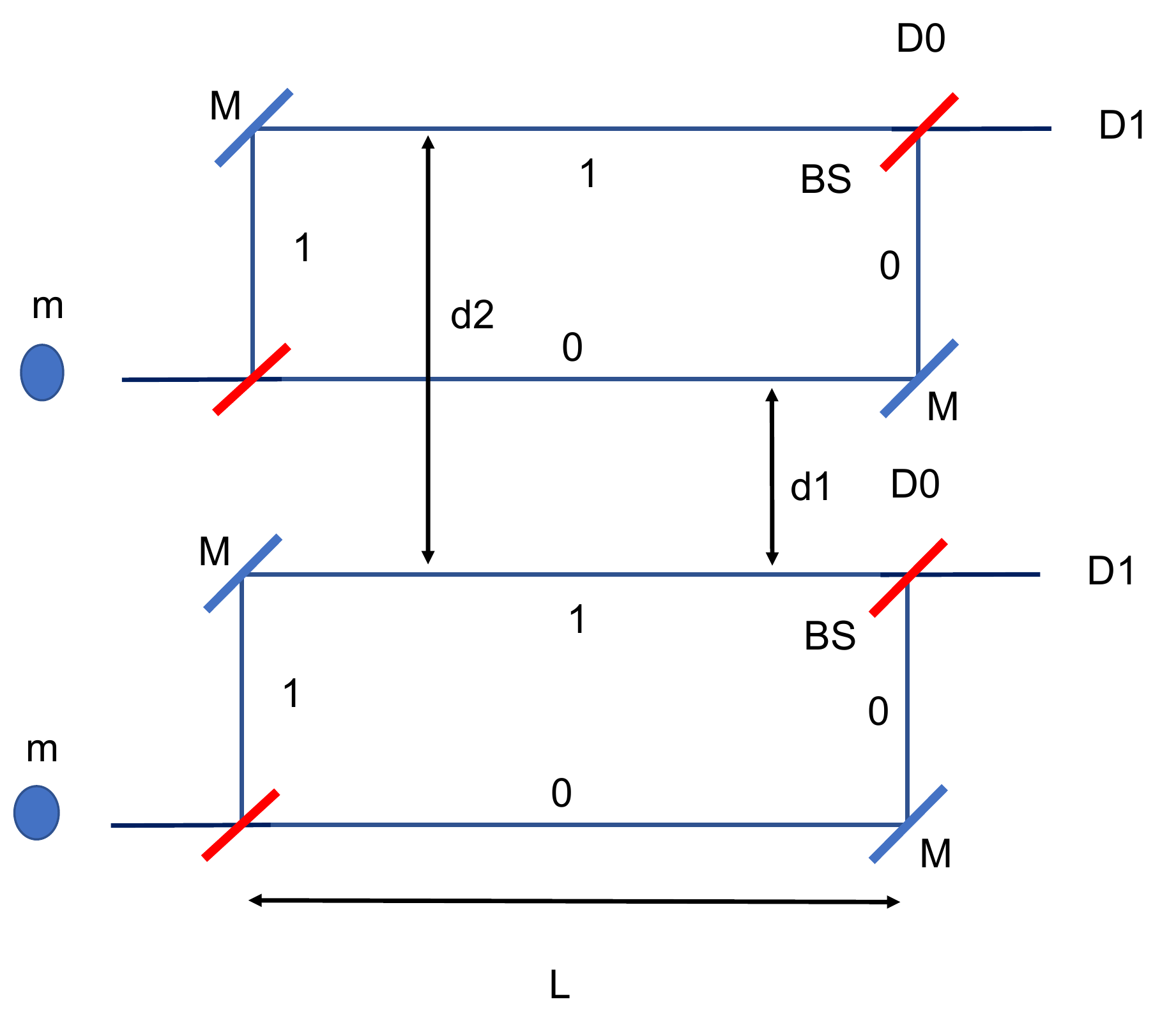} 
	\caption{Entanglement-based witness of quantum gravity with two equal masses. Each mass $m$ individually undergoes Mach-Zehnder-type interference, and interacts with the other mass via gravity. BS indicates a beam splitter; M indicates a mirror; Di with $i=0,1$ indicates the detector on path $i$. $L$ is the length of the lower arms of each interferometer. The distance between the lower arms of the two interferometers is $d_1$ and the distance between the upper arm of one interferometer and the lower arm of the other interferometer is $d_2$.
See full description in the text.}
	
\end{figure}

Since the masses on different paths interact via the gravitational field, the state of the composite system becomes, before they enter their respective final beam-splitters: 
\begin{eqnarray}
\frac{1}{2}\ket{0}\left(\ket{0}+\exp{(i\phi_1)}\ket{1}\right )+ \nonumber \\ 
\frac{1}{2}\exp({i\phi_1})\ket{1}\left(\ket{0}+\exp{(i(\Delta{\phi}))}\ket{1}\right )\;.
\end{eqnarray}

\noindent where $\phi_1$ and $\phi_2$ are the relative phases acquired by the masses due to the gravitational potential generated when they are, respectively, at distance $d_1$ and $d_2$ from one another; $\Delta\phi=\phi_2-\phi_1$ is their difference. We suppose that the gravitational interaction of the masses on the two most distant arms is negligible. Supposing that the dominant contribution to the gravitational interaction is Newtonian, and that the general-relativistic contributions are negligible, the value of the phase is $\phi_i=\frac{m^2G}{\hbar d_i}\Delta t$; where $G$ is the gravitational coupling constant; $\Delta t=\frac{L}{v}$ is the time spent by each mass on the horizontal arm of the interferometer, of length $L$; and $v$ is their velocity. However, the conclusion as for the quantisation of the gravitational field would be the same no matter what type of field mediates the entanglement. It is remarkable, thought, that even the Newtonian contribution already demonstrates the quantum nature of gravity.

Depending on the particles' mass, the distance between the two interferometers and the length of the arms, the above state is entangled to a different degree. The mutual interaction of each of the masses acts as a measurement of which-path they are on: depending on the phases, that interaction may completely destroy the interference effect of each mass, showing maximal entanglement. 

In each of the interferometers, the probabilities $p_{\alpha}$ for the mass to emerge on path $\alpha=0,1$ are:

\begin{equation}
p_0=\frac{1}{2}\left(\cos^2{\frac{\phi_1}{2}}+\cos^2{\frac{\Delta \phi}{2}}\right)\;,\\
p_1=\frac{1}{2}\left(\sin^2{\frac{\phi_1}{2}}+\sin^2{\frac{\Delta \phi}{2}}\right)
\end{equation}

There are two extreme regimes. One is when the two masses are maximally entangled by the action of the gravitational field, in which case $p_0=p_1=\frac{1}{2}$. This happens when $\phi_1=2n\pi$, $\Delta\phi={\pi}$ for some integer $n$. The other extreme is when the two masses are not entangled and undergo, separately, an ordinary interference experiment: that happens when $\phi_1=\Delta\phi=2n\pi$. In this case, each mass emerges on path $0$ of the interferometer. For a fixed mass, by varying the arms' distance or their length, it is in principle possible to interpolate between those two cases, thus demonstrating all degrees of entanglement, ranging from no entanglement to maximum. By our argument, this entanglement is a witness that the gravitational field that mediated the interaction must be quantum.  

Feasibility considerations suggest that the experiment can be realised with existing technologies. The two masses could be massive molecules, as in \cite{ARN}, two split Bose-condensates \cite{ASE}, or two nano-mechanical oscillators \cite{NANO}. For example, two coupled nano-mechanical oscillators of mass $10^{-12}$ kg interacting for a $\Delta t=10^{-6}$ s would achieve the extreme phase shifts, over distances $d\approx 10^{-6}$m. The main difficulty with the experiment is that any other effects on the masses must be made smaller than the gravitational interaction. But there are other important subtleties which we proceed to discuss. 

Failing to detect entanglement would not imply that gravity is not quantum. It could be that our model is inadequate, because the interaction between the field and the masses contains higher order correlations. In our analysis we have added up the phases linearly (as one would do for the electromagnetic case); but this need not apply to gravity \cite{STAMP}. It could also be that the field is quantum, as defined above, but it is a decoherent channel -- as discussed in \cite{MIL}. In this case, the field would have non-commuting degrees of freedom, but they would be so decohering that the channel could never transmit any entanglement (i.e. it would be 'entanglement breaking'). The type of decoherence depends on the particular implementation for our proposed experiment. For example, for nano-mechanical oscillators, the time-scale for decoherence can be expected to be in the range of $\mu$s to ms, \cite{nano1, nano2}, therefore any time interval below $\mu$s would be acceptable. 

If the experiment succeeded in detecting entanglement between the masses, one would have to make sure that it is really generated mainly by gravity. There could be other sources of interaction between the masses that would lead to entanglement -- for example the Van der Waals forces or other electromagnetic interactions. If they were much stronger than gravity, we could not conclude anything about the quantisation of gravity. However, if they were comparable, it might still be possible to isolate the characteristic $\frac{1}{r}$ behaviour of the gravitational potential in the phase (as opposed for example to the Lennard-Jones $\frac{1}{r^6}$); in which case the proposed scheme might still apply.  Finally, detecting entanglement, even when it is generated only through gravity, does not indicate which model of quantised gravity one must adopt. Whatever the model, the conclusion is that the field must have at least two non-commuting observables. What these observables are remains an open question (see supplement for further speculations).

It is illuminating to analyse how the two field observables entangle the two masses, if the field indeed is quantum. First, each mass becomes entangled with the field, interacting with one of the field's observable. This is exactly what Feynman had in mind for a single mass. Then the phases are induced into each of the four massive configurations (as in Fig. 1) via the interaction between the masses and the other observable of the field. Finally, the field becomes disentangled from the masses as they emerge from the interferometer (again as would happen in Feynman's experiment with a single mass). 

Remarkably, our proposal works even if the entanglement between the masses and the field is very small; because it relies on detecting only the entanglement between the positions of the two masses. In fact, the entanglement between the field and the masses may never even be detectable in practice just like a spontaneous emission of a graviton \cite{DYS}. Still, it is sufficient to generate the phases leading to entanglement between the masses.   

Since our proposal does not rely on any specific dynamics, it could also be applied to prove the quantisation in different scenarios. For example, one could think of variants of the Aharonov-Bohm (AB) effect, where the solenoid interacting with a quantum charge is replaced by a small current caused by another quantum charge, generalising recent proposals to explain the AB effect purely in terms of quantum entanglement \cite{VAI}. 


\noindent The key next steps are to identify the best physical implementation of our proposal; to provide a detailed analysis of all the relevant effects that could compete with gravity; and to perform a comparative study of how this experiment would be modelled in different approaches to quantum gravity. 

\section{Supplementary information}

We now discuss a simple example to illustrate how the quantised gravitational field could mediate the entanglement between the two masses. This is by no means a proposal for how to quantise gravity -- its sole purpose is to explain how the gravitational field's complementary observables establish the entanglement between the spatial degrees of freedom of the masses. It will also allow us to speculate about the meaning of these complementary observables in the gravitational field; and show that no matter how little entanglement there is between the field and the masses, the two masses can still get (even maximally) entangled with one another via the gravitational interaction.

Consider a simple model where the gravitational field $C$ is treated as a quantum harmonic oscillator (or, more accurately, a collection of them), in conformity with the traditional linearised models of quantum gravity, such as \cite{BOG}. The two masses $Q_1$ and $Q_2$ are modelled as two qubits -- whose z-component represents a discretised position of each mass. (In our earlier discussion, the position represents one of two arms of a Mach-Zehnder interferometer, but it could be more general.) Its eigenstates $\ket{a}$ where $a\in \{0,1\}$ represent the situation where the mass is in a definite position $a$; the state $\ket{ab}$ describes the situation where the first mass is in position $a$ and the second in position $b$. 

To analyse the formation of entanglement between the two masses, we only need to account for how the relative phases in the quantum superpositions of masses are established during the double interference experiment (as in figure 2). This happens via first entangling the masses to the field, then generating the phases, and finally disentangling the field from the masses. Following the logic of our proof, we describe the process by giving the quantum states of the composite system of the two masses and the field. How to realise a superposition of each mass (i.e. how to implement the first beam splitter) depends on the details of the particular implementation. The same goes for their final interference (i.e. the second beam splitter).
Immediately after the action of the first beam splitter, the state of the two masses and the field is
$$\ket{\phi_0}=\frac{1}{2}\sum_{a,b\in\{0,1\}}\ket{ab}\ket{\alpha}$$
where $\ket{\alpha}={\rm e}^{-\frac{1}{2}|\alpha|^2}\exp(\alpha (a^{\dagger}-a))\ket{0}$ is a coherent state which
represents many spatial modes of gravity - possibly a continuum - and $a, a^{\dagger}$ are the bosonic creation and annihilation operators. We are using a coherent state because this is the best quantum
approximation of the classical field. 
The two masses and the field then evolve into the state $$\ket{\phi_{E1}}=\frac{1}{2}\sum_{a,b\in\{0,1\}}\ket{ab}\ket{\alpha_{a,b}}$$ where $
\ket{\alpha_{a,b}}=\ket{\alpha+\sqrt{\xi_{ab}}}=D(\xi_{a,b})\ket{\alpha}\;$ and we have defined the displacement operator as $D(\xi_{a,b})=\exp{(\sqrt{\xi_{a,b}}(a^{\dagger}-a))}$ with $\sqrt{\xi_{ab}}$ being a shift that depends on the coupling between the field and the masses, that brings about the desired phase-shift $\phi_{a,b}$ at the end (see below for more details). For simplicity, we assume that $\xi_{ab}$ is real. We assume that establishing the entanglement
between the field and the masses takes place on time-scales much faster than the process that transfers the phase $\phi_{a,b}$ back from the field to the masses, evolving their composite system to the state $$\ket{\phi_{E2}}=\frac{1}{2}\sum_{a,b\in\{0,1\}}\exp{(i\phi_{a,b})}\ket{ab}\ket{\alpha_{a,b}}\;.$$ Finally, the interaction between the field and the masses brings the field back to its original state and the masses remain entangled (to the degree depending on the phase): $$\frac{1}{2}\sum_{a,b\in\{0,1\}}\exp{(i\phi_{a,b})}\ket{ab}\ket{\alpha}\;.$$

The key fact is that the above process relies on two complementary observables of the field. This can be seen by noticing that the observables $\frac{1}{2i}(a-a^{\dagger})$ and $a^{\dagger}a$ are needed to express the intermediate states $\ket{\phi_{E1}}$ and $\ket{\phi_{E2}}$ in terms of the initial state $\ket{\phi_0}$. Indeed, supposing for simplicity that initially $\ket{\alpha}=\ket{0}$, $\ket{\phi_{E1}}=\sum_{a,b\in\{0,1\}}P_{ab}\otimes D(\xi_{ab})\ket{\phi_0}$ and $\ket{\phi_{E2}}=\exp(w(a^{\dagger}a)) \ket{\phi_{E1}}$, where we have defined the projectors $P_{ab}=P_a\otimes P_b$, with $P_{0,1}=\frac{(id\pm\sigma_z)}{2}$ being the projector operator for the location of each mass; and $w$ is some real number with the property that $w\xi_{a,b}=\phi_{a,b}$.

The values $w$ and $\xi_{ab}$ depend on the particular model of interaction between gravity and the masses. It is important to notice that, whatever the model, the amount of entanglement between the masses and the field depends on the numbers $\xi_{a,b}$ only (and not on $w$). 
The entanglement between the field and masses is given by the reduced entropy of the masses. In this regime, where the field and the masses are weakly entangled, this entropy is well approximated by the linear entropy ($S_L=1-{\rm Tr(\rho_{Q_1,Q_2}^2)}$, where $\rho_{Q_1,Q_2}$ is the reduced state of the two masses). The magnitude of the reduced entropy is given by one minus the overlap between the two gravitational states squared as in:
\begin{equation}
1- |\langle\alpha_{ab}|\alpha\rangle|^2 = 1- \exp{(-\xi_{ab})} \approx \xi_{ab}\; .
\end{equation}
This quantity could be small (compared to unity), but that does not affect the efficacy of the proposed indirect witness. For instance, assuming one is dealing with Newtonian gravity, as we did in our discussion, one would have:
$$
\phi_{ab}=w\alpha_{ab}=\frac{Gm^2}{\hbar d_{ab}}\Delta t=\left (\frac{m}{m_P}\right )^2\frac{c}{d_{ab}}\Delta t\;.
$$
where $\Delta t$ is the interaction time between the two masses, $m_{P}$ is Planck's mass and $d_{ab}$ is the distance between the position $a$ of the first mass and position $b$ of the second mass. One can identify $\alpha_{ab}=\left (\frac{m}{m_P}\right )^2$. Thus, assuming that $m=10^{-12}$kg, the degree of entanglement between the masses and the field could be $10^{-12}$ (depending on the details of the model, this entanglement could in fact be much smaller).

Entanglement being very small might be linked to the fact that the spontaneous emission of gravitons is undetectable.  In other words, the phenomenon of entangling the masses, though mediated via gravitons, does not rely on gravitons being detectable. To see this clearly, we can identify what the complementary observables of the field would be in a linearised model of quantum gravity, where $a$ and $a^{\dagger}$ would then be the annihilation and creation operators for gravitons. In the linear regime, the Hamiltonian for the free gravitational field contains the graviton number operator \cite{BOG}
$$
H_{Free}=\sum_{\sigma}\int d^{3}k \hbar \omega_k \left [a_{k\sigma}^{\dagger}a_{k\sigma}+ \frac{1}{2} \right].
$$
In the above we have included two polarization degrees of freedom of the field using the label $\sigma$, while $\omega_k$ and $k$ represent the frequency and wavenumber of the mode respectively.
The corresponding linearised gravity-matter interaction Hamiltonian contains the complementary observable $\frac{1}{\sqrt{2}}(a+a^{\dagger})$. It is obtained from the general linearised Hamiltonian:
\begin{equation}
H^G_{int} =- \frac{1}{2} h_{\mu\nu} T^{\mu\nu}
\end{equation}
where $T^{\mu\nu}$ is the stress-energy tensor and $h_{\mu\nu}$ is the perturbation of the metric tensor $g_{\mu\nu}$ away from the flat (Minkowski) spacetime. In our experiment, the masses are non-relativistic and the stress-energy tensor would simplify to $T_{00} = m$. 
The quantized gravitational field is then written in terms of the graviton creation and annihilation operators as:
\begin{equation}
h_{\mu\nu} \propto \sum_{\sigma} \int \frac{d^3k}{\sqrt{\omega_k}} \{ a(k,\sigma)\epsilon_{\mu\nu} (k,\sigma) e^{ik_\lambda x^\lambda} + h.c.\}
\end{equation}
where $\epsilon_{\mu\nu}$ is the polarization tensor and we are using the Einstein's convention of summation for the phase. Therefore even the simplest linearised version of quantum gravity would contain the two complementary observables one needs to generate entanglement in our experiment. 

It would be interesting to analyse the phenomenon that leads to the generation of entanglement between the masses and the field comparing different quantum gravity models; each one of them might lead to different predictions. Also, our analysis implies that semi-classical versions of quantum gravity (such as the one in quantum field theory in curved space-time) would not lead to any spatial entanglement between the two masses. We will address these issues in a forthcoming paper.

\textit{Note added}: After the completion of our work we became aware of a related parallel independent work \cite{BOS}.

\textit{Acknowledgments}: 
The Authors thank two anonymous referees for helpful comments. CM's research was supported by the Templeton World Charity Foundation and by the Eutopia Foundation. VV thanks the Oxford Martin School, the John Templeton Foundation, the EPSRC (UK). This research is also supported by the National Research Foundation, Prime Minister's Office, Singapore, under its Competitive Research Programme (CRP Award No. NRF- CRP14-2014-02) and administered by Centre for Quantum Technologies, National University of Singapore.

\end{document}